\documentclass[twocolumn]{aastex631}

\usepackage{graphicx} 
\usepackage{comment}
\usepackage{amsmath}
\usepackage{multirow}
\usepackage{booktabs}
\usepackage{enumitem}
\usepackage{color}
\newcommand\s{{\rm~s}}

\newcommand\kev{{\rm~keV}}

\newcommand\cm{{\rm~cm}}

\newcommand\hst{{\it HST}}

\newcommand\astrosat{\textit{AstroSat }}

\graphicspath{{./}{Figures/}}

\begin{document}

\title{Investigating Accretion Disk–Corona in Seyfert 1 galaxies: A UV/X-ray Spectral Study of Mrk~813 and RBS~688}

\author{Piyali Ganguly}
\affiliation{Inter-University Centre for Astronomy and Astrophysics (IUCAA), Post Bag 4, Ganeshkhind, Pune 411007, India\\}

\author{Gulab C. Dewangan}
\affiliation{Inter-University Centre for Astronomy and Astrophysics (IUCAA), Post Bag 4, Ganeshkhind, Pune 411007, India\\}

\begin{abstract}

We present a broadband UV/X-ray spectral study of two Seyfert 1 galaxies, Mrk~813 and RBS~688, primarily based on \astrosat{} observations.  These active galactic nuclei host relatively large super-massive black holes ($M_{BH} \sim 10^8 - 10^9M_{\odot}$), suffer negligible internal extinction/absorption, and are well suited for probing the inner regions of their accretion disks using far UV and soft X-ray spectra. In the case of Mrk~813, the \astrosat{} and \hst{} far UV spectra are steeper than those expected from a standard accretion disk; the deficit of emission at shorter wavelengths suggests a truncated accretion disk with an inner radius $r_{in} \sim 70r_g$. Joint UV/X-ray broadband spectral modelling with {\sc fagnsed} and {\sc relagn} models suggests that the apparent truncation in Mrk~813 is most likely due to the presence of a warm Comptonising disk in the inner regions that is responsible for the observed soft X-ray excess emission. RBS~688 lacks the soft X-ray excess emission, and the UV data are entirely consistent with a standard disk that appears to extend very close to the innermost stable circular orbit. Our study suggests the formation of the warm, optically-thick Comptonising corona in the innermost disk regions at higher Eddington fraction. 
\end{abstract}


\section{Introduction} \label{sec:intro}
Active Galactic Nuclei (AGN) are one of the most luminous objects in the universe with their bolometric luminosity in the range of $\sim10^{40} - 10^{48}{\textrm~ergs~s^{-1}}$. The source of such an enormous amount of energy is believed to be the accretion disk around the supermassive black holes at the center of these galaxies. These central black holes, having masses in the range of $10^5-10^{10} M_{\odot}$, accrete the gas around them in the form of a disk, and the gravitational energy of the in-falling matter is radiated out in the form of electromagnetic radiation throughout a wide range of frequency, from radio to $\gamma$-rays.

Observationally, the Optical/UV continuum or the Big Blue Bump (BBB) extends from near-infrared at $\sim 1 \mu$ to extreme ultraviolet at $\sim 1000$ \AA. The BBB, accounting for more than half of the bolometric luminosity in type-1 AGN, is thought to be the direct signature of the accretion flow \citep{1983ApJ...268..582M, 1987ApJ...321..305C, 1994ApJS...95....1E, Koratkar_1999, Shang_2005}. However, the nature of the accretion disks has not been clearly understood, as in the case of  X-ray binaries, where optically thick and geometrically thin standard accretion disks \citep{1973A&A....24..337S} have been successful (\citep[see e.g.,][]{2006ApJ...647..525D, 2006ARA&A..44...49R}. The observations of relativistically-broadened iron K$\alpha$ lines and the related X-ray reflection emission from several AGN have indirectly revealed the presence of optically thick material in the form of disks around the SMBHs \citep[see e.g.,][]{2000PASP..112.1145F, 2007ARA&A..45..441M}. However, these observations can neither probe the intrinsic emission from the accretion disks nor the radial dependence of the disk temperature, which is essential to unravel the nature of the disk. Contamination from the host galaxy, numerous emission lines from broad and narrow line regions, intrinsic extinction, lack of high-quality ultraviolet data, etc., further introduce complexity in unravelling the intrinsic disk continuum emission. 

Some of the above complexities can be avoided by carefully selecting AGN with low disk temperatures and observing them with instruments sensitive in the far UV band, such as the Ultra-Violet Imaging Telescope (UVIT) onboard \astrosat{}. The standard accretion disk theory predicts that the temperature of the inner accretion disk decreases with black hole mass as $T \propto M_{BH}^{-1/4}$. With increasing masses of the central black holes, the peak of the emitted spectrum shifts towards longer wavelengths, and the peak falls in the far UV band for $M_{BH} \gtrsim 10^9 M_\odot$. It should also be noted that the host galaxy contamination and the intrinsic extinction are minimised in quasars, which are point-like, unabsorbed type 1 AGN. A significant fraction of type-1 AGN also exhibits soft X-ray excess emission below 2~KeV, and depending on the model of its origin, this component can also complicate the measurement of the intrinsic disk emission.

\begin{deluxetable*}{cccccc}
\tablecaption{Basic parameters of Mrk~813 and RBS~688.
\label{Table:src_table}}
\tablehead{\colhead{Source} & \colhead{RA} & \colhead{DEC}  & \colhead{Redshift} &  \colhead{Black hole mass}   & \colhead{Galactic $N_H$ }\\
\colhead{} & \colhead{(hh:mm::ss)} & \colhead{(dd:mm:ss)} & \colhead{}  & \colhead{$M_{BH}(M_\odot$)}   & \colhead{($ 10^{20} cm^{-2}$)}}
\startdata
Mrk 813 &  14:27:25.05 & +19:49:52.27  & 0.111 & $9.0 \times 10^7$ & 2.29\\
RBS 688 &  08:01:31.97 & +47:36:16.06 & 0.157 & $9.55\times 10^8$ & 3.93\\
\enddata
\end{deluxetable*}

In this paper, we study the far UV and X-ray emission from two type~1 AGN, namely Mrk~813 and RBS~688, using the simultaneous observations performed with the UVIT and the Soft X-ray Telescope (SXT), onboard  \astrosat \citep{2014SPIE.9144E..1SS}. 
 
We selected our targets from SPIDERS (Spectroscopic Identification of eROSITA sources) catalogue  \citep{2019A&A...625A.123C} with a black hole mass above $10^{8.5} M_\odot$. For each source in SPIDERS catalogue the black hole mass was obtained using single-epoch mass estimation method with H$\beta$ and/or Mg~II line. The Black hole mass thus measured for Mrk~813 and RBS~688 are respectively $9.8\times 10^8 M_\odot$ and $9.5\times 10^8 M_\odot$. 
 
However in a very recent work by \cite{2024ApJS..272...29Z} under the MAHA (Monitoring AGN with H$\beta$ Asymmetry) campaign, the mass of the central black hole of Mrk~813 using reverberation mapping technique is found to be $17.0_{-5.8}^{+7.1} \times 10^7 M_\odot$ and $9.0_{-3.2}^{+3.9} \times 10^7 M_\odot$ calculated from FWHM and $\sigma_{line} $ of the ${H\beta}$ line, respectively. In this paper, we have used $9 \times 10^7 M_\odot$ for the mass of the central black hole of Mrk~813. More details about the targets are mentioned in Table~\ref{Table:src_table}. With simultaneous observations in both far UV and X-ray bands, both of these sources present us an excellent opportunity to probe the innermost regions of the accretion disk around SMBHs at the moderate to high mass end of the population. We describe details of our observation and data reduction  in Section~\ref{sect:data}, followed by the spectral analysis in Section \ref{sect:analysis}. we discuss our findings  in Section \ref{sect:result}.

\section{Observations and Data Reduction}\label{sect:data}
In this paper, we have primarily used UV and X-ray data acquired with the \astrosat{} observatory \citep{2014SPIE.9144E..1SS}. \astrosat{} is a multi-wavelength space observatory launched by the Indian Space Research Organisation (ISRO). It houses four different payloads that operate simultaneously. These payloads are the Ultra-Violet Imaging Telescope (UVIT; \citealt{2017AJ....154..128T, 2020AJ....159..158T}), the Soft X-ray Telescope (SXT; \citealt{2016SPIE.9905E..1ES, 2017JApA...38...29S}), the Cadmium-Zinc-Telluride Imager (CZTI; \citealt{2016SPIE.9905E..1GV}) and the Large Area X-ray Proportional Counters (LAXPC; \citealt{2016SPIE.9905E..1DY, 2017JApA...38...30A, 2017ApJS..231...10A}). In this paper, we have utilised the simultaneous data acquired with the UVIT and SXT. The data from the other two instruments, CZTI and LAXPC, are dominated by background and are not useful.

The UVIT consists of two co-aligned telescopes that collectively acquire images in three channels: FUV (1200-1800 \AA), NUV (2000-3000 \AA) and VIS (3200-5500 \AA). Only the data acquired in the FUV and NUV bands are used for scientific observations, and the visible band data are mainly used for tracking the satellite's pointing. 
The NUV channel is no longer in operation since March 2018 \citep{ghosh2021orbit}.
The FUV channel hosts a number of broadband filters and two slitless gratings. It operates in the photon-counting mode at a rate of 28 frames per second in full-window mode.

\begin{deluxetable*}{cccccc}
\tablecaption{\astrosat{} observations of Mrk~813 and RBS~866.
\label{Table:obs_table}}
\tablehead{\colhead{Source} & \colhead{Obs. ID} & \colhead{Date of obs.}  & \colhead{Instrument} &  \colhead{Exposure time}   & \colhead{count rate}\\
\colhead{} & \colhead{} & \colhead{} & \colhead{}  & \colhead{(ks)}   & \colhead{(counts/sec)}}
\startdata
Mrk 813 & A11\_106T01\_9000005102 & 2022-05-02 & UVIT/FUV G1 & 8.4 & $2.27 \pm 0.02$\\
& & & SXT & 28.2 & $0.100 \pm 0.002$ \\
RBS 688 & A11\_106T02\_9000005344 & 2022-20-09 & UVIT/FUV G1 & 4.27 & $2.02 \pm 0.02$\\
& & & SXT & 45.8 & $0.056 \pm 0.002$\\
\enddata
\end{deluxetable*}

The SXT is sensitive to soft X-rays in the energy range of 0.3-8.0 keV with an energy resolution of $\ sim90$eV at 1.5~keV and $\sim136$eV at 5.9~keV. SXT has a tubular structure that houses the X-ray reflecting mirrors and other components. The SXT houses a charged-coupled device (CCD) placed at the focus of the mirrors. The mirror assembly of X-ray Telescope in SXT consists of a set of coaxial and con-focal shells of conical mirrors approximating paraboloidal and hyperboloidal shapes and arranged behind each other following a geometrical arrangement called Wolter-I optics. The focal length of the telescope is 2~m and the effective area is 90 $\text{cm}^2$ at 1.5 KeV. The telescope has a circular field of view with a diameter of $40^{\prime}$. The on-axis FWHM of the point spread function (PSF) in the focal plane is $2^{\prime}$ and the Half-Power Diameter is $11^{\prime}$. The Focal Plane camera assembly houses a CCD along with four individual calibration sources for in-flight calibration at energies of $\approx 5.9$ KeV. These calibration sources illuminate the corners of the CCD outside the field of view.   

\begin{figure*}
    \centering
    \gridline{\fig{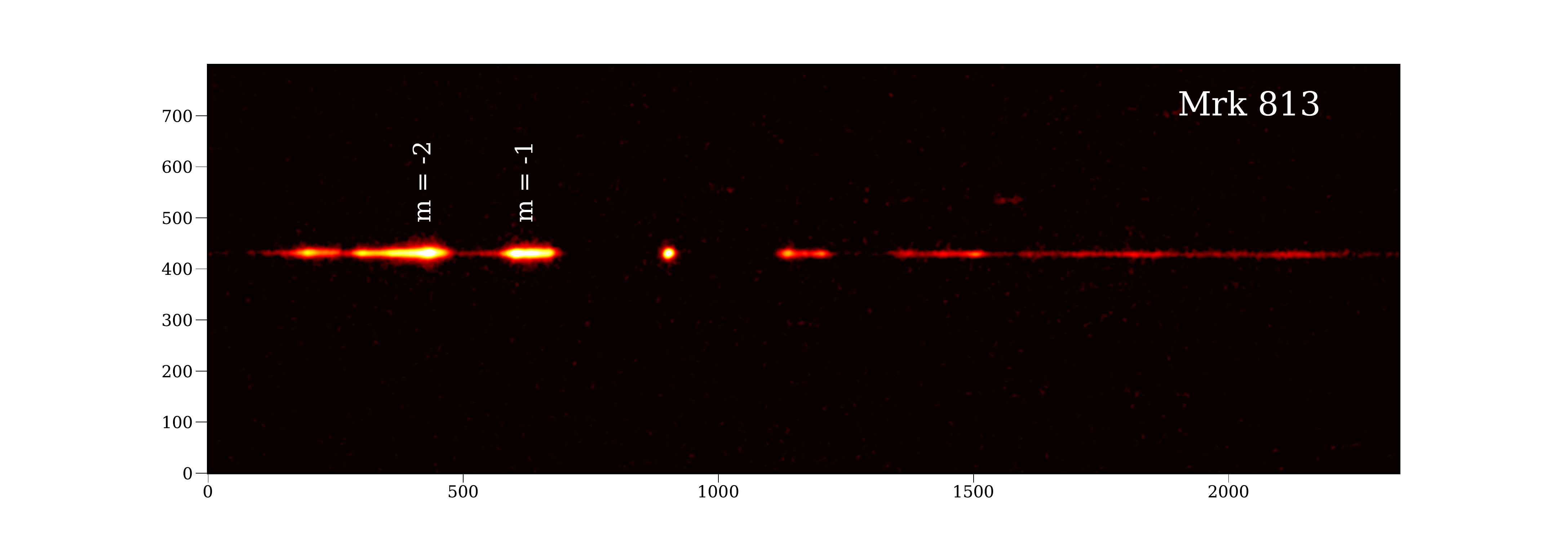}{\textwidth}{}}
    \gridline{\fig{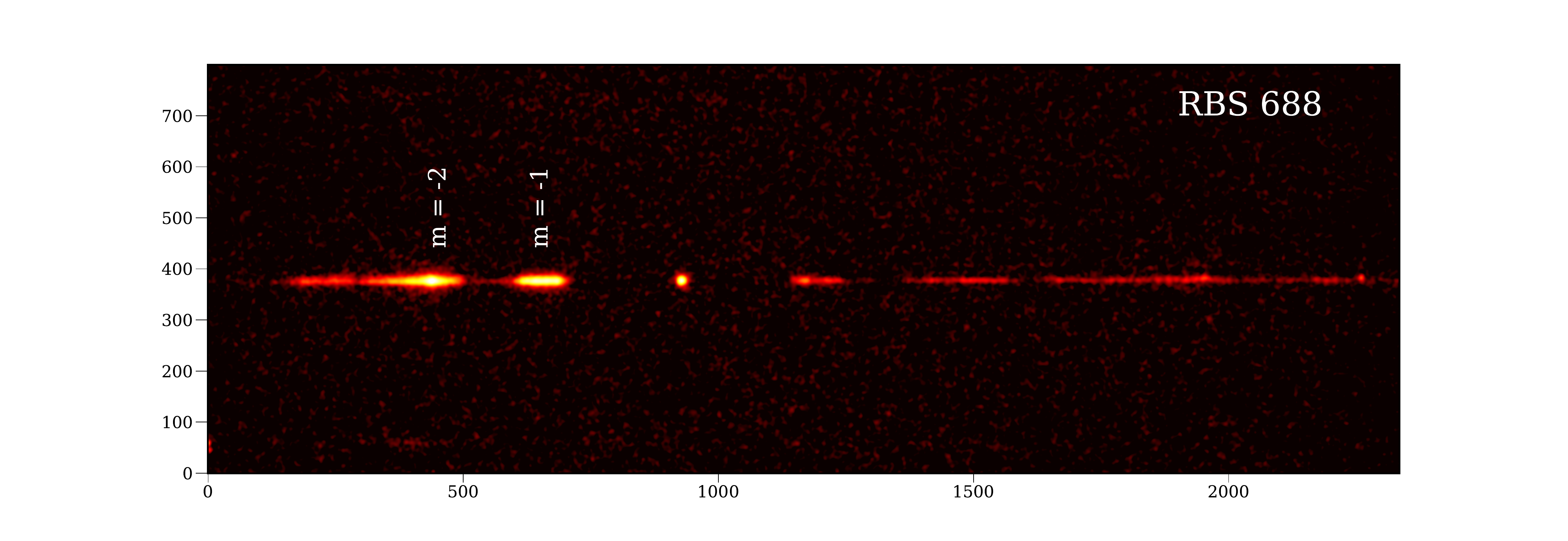}{\textwidth}{}}
    \caption{The UVIT/FUV-G1 slitless grating images of Mrk~813 (upper panel) and RBS~688 (lower panel).}
    \label{fig:UVIT_grating_images}
\end{figure*}

For both the targets we obtained Level-1 FUV grating data from the \astrosat{} data archive\footnote{\url{https://astrobrowse.issdc.gov.in/astro_archive/archive/Home.jsp}}. We processed the FUV grating data with {\sc CCDLAB} pipeline \citep{2017PASP..129k5002P} which is a dedicated software developed for the UVIT data processing. From the level-1 data, we generated cleaned images for each orbit after correcting for the drift in the satellite pointing direction and merged all orbit-wise images to produce a single image. The exposure time and other details of each observation are listed in Table~\ref{Table:obs_table}. We show the 2D grating images for both of our targets in Figure~\ref{fig:UVIT_grating_images}. We used the UVITTools.jl package\footnote{\url{https://github.com/gulabd/UVITTools.jl}} described in \cite{2021JApA...42...49D}, and extracted the 1D spectrum for each of our targets using a cross-dispersion width of 50 pixels. Since the order with maximum efficiency for the FUV grating is the -2 order, we used the grating spectra from that order only. To be able to model both the UV and X-ray spectra jointly using X-ray spectral fitting packages, we converted the count spectrum to the equivalent pulse height amplitude (PHA) spectrum following the prescription in \cite{2021JApA...42...49D}. We also created background PHA spectra using source-free regions in the 2D images. We used the grating response files available within the UVITTools.jl package.

 \begin{figure*}
     \centering
     \gridline{\fig{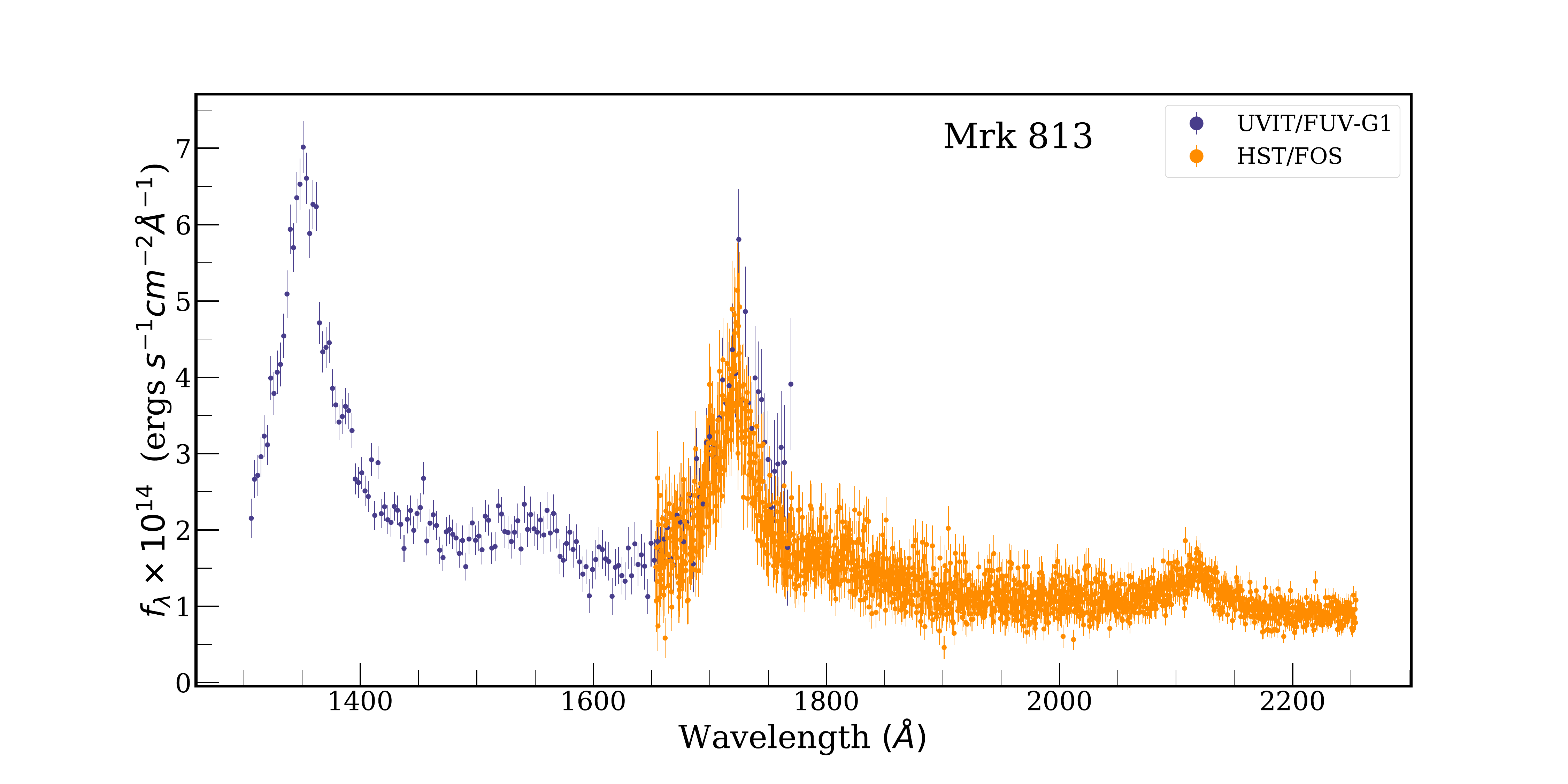}{\textwidth}{}}
     \gridline{\fig{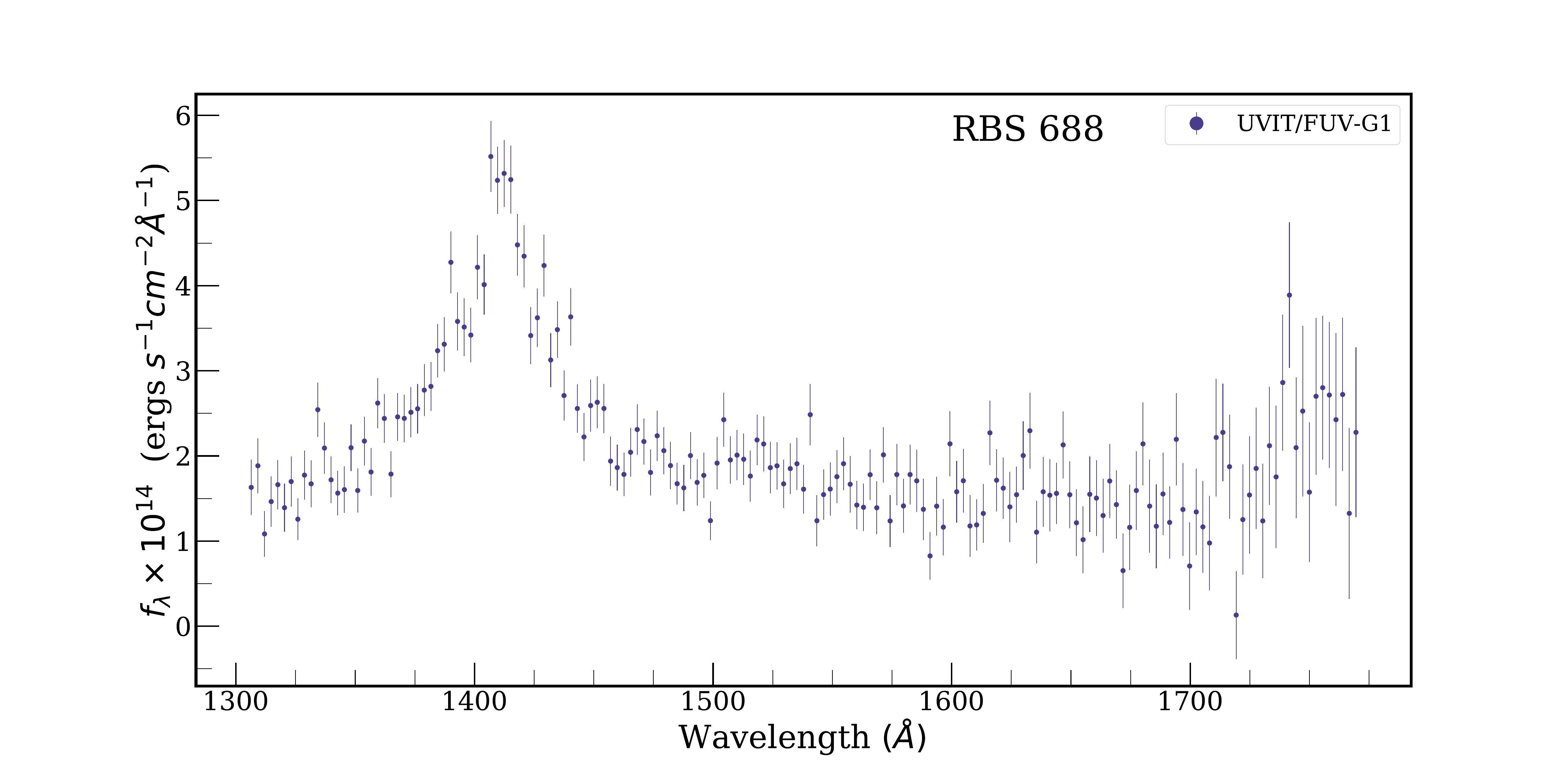}{\textwidth}{}}
     \caption{The UVIT/FUV-G1 and \hst/FOS spectra of Mrk~813 (upper panel) and the FUV-G1 spectrum of RBS~688 (lower panel).
     \label{fig:UV_spectrum}}
 \end{figure*}

For the SXT observations, we downloaded the orbit-wise level-2 SXT data, which were processed using the SXT pipeline (version 1.4b) by the Payload Operation Centre.
We then merged the processed orbit-wise SXT event lists using the Julia merger tool SXTMerger.jl\footnote{\url{https://github.com/gulabd/SXTMerger.jl}}.  We used the HEASOFT tool {\sc xselect} to generate the PHA spectral data using a circular region of radius $15^{\prime}$ around the source position. Due to the large PSF wings with half power diameter of $11^{\prime}$, there is no reasonably sized source-free region available in the CCD frames. Therefore, we used the standard background spectrum\footnote{available at \url{https://www.tifr.res.in/astrosat_sxt/dataanalysis.html}\label{footnote1}} provided by the SXT team. We used the most recent re-distribution matrix file\footref{footnote1} and the ancillary response file\footref{footnote1} available for SXT. Using {\sc ftgrppha} we grouped our SXT data to have at least 25 counts per bin.

\subsection{\hst{}/FOS spectrum}
To support our UVIT spectral data, we searched for the availability of \hst{} spectra of our sources. We found the \hst/FOS spectrum of Mrk~813 in the archive\footnote{\url{https://hea-www.harvard.edu/~pgreen/HRCULES.html}} \citep{2002ApJS..143..257K}, but RBS~688 has not been observed with the \hst{}. We obtained the flux-calibrated \hst{}/FOS spectrum of Mrk~813, and converted to XSPEC-compatible spectral data using the {\sc ftflx2xsp} tool available as part of the HEASOFT package. The FOS spectrum covers the $ 1590-2312$\AA\, band, with a spectral resolution (FWHM) $\sim 1.41$\AA. We used the FOS spectrum in the 1650--2254~\AA range as the data below 1650~\AA have large errors. We show the UVIT/FUV-G1 and \hst{}/FOS spectra in Figure~\ref{fig:UV_spectrum}. We note that the FOS and the FUV-G1 spectra agree very well in the overlapping band. Hence, these data are suitable for simultaneous spectral fitting.

\section{Analysis}\label{sect:analysis}
 We employed  XSPEC (version 12.12.1) \citep{1996ASPC..101...17A} for our spectral analysis. We used the $\chi^2$ statistics for goodness of fit, and quoted the errors at the $90\%$ confidence level, unless otherwise specified. First, we analysed UV spectral data, followed by X-ray data, and finally, we performed joint spectral analysis of UV and X-ray data.

\subsection{UV Spectral Analysis}\label{subsection:UV_analysis}
 The observed spectra of active galactic nuclei can include significant contributions from other components, such as the broad and narrow line regions and the host galaxy, in addition to the continuum emission from the accretion disk,
The observed spectra are also affected by the Galactic and possibly internal reddening. To account for the Galactic reddening, we have used the {\sc XSPEC} model {\sc redden}. This model uses the \cite{1989ApJ...345..245C} extinction law with $R_V$ = 3.1, and has only one parameter -- the color excess E(B-V) which we calculated from the Galactic $N_H$ using the relation \citep{2009MNRAS.400.2050G},

\begin{equation}
N_H = (6.86 \pm 0.27) \times 10^{21} E(B-V)
\end{equation}
We obtained the Galactic $N_H$ values, listed in Table~\ref{Table:src_table}, from the $N_H$ calculator tool available at the HEASARC website\footnote{\url{https://heasarc.gsfc.nasa.gov/cgi-bin/Tools/w3nh/w3nh.pl}}.
The calculated E(B-V) values for Mrk~813 and RBS~688 are 0.033 and 0.057, respectively.

We first determined the contribution of emission lines. For this purpose,  we used a simple power-law model ({\sc zpowerlw}) for the continuum and Gaussian line profiles for the broad and narrow lines, all modified by the Galactic reddening. This model provided a satisfactory fit to the FUV-G1$+$FOS spectra of Mrk~813 and FUV-G1 spectrum of RBS~688. We also tested for internal reddening using the {\sc XSPEC} model {\sc  intr\_ext} implemented by \citet{2024ApJ...975...73K} following the empirical extinction relation given by \cite{10.1111/j.1365-2966.2004.07590.x}. We found that for both AGN, the fit statistics did not improve upon using an intrinsic reddening component, and the continuum parameters did not change. So, we concluded that this component was not required for both the AGN and was not used in any subsequent analysis. We list the best-fit emission line parameters for both the AGN in Table~\ref{Table:emission_lines}.

In the case of Mrk~813, we found a total of 8 lines, as mentioned in Table~\ref{Table:emission_lines}; three of these lines (two Ly$\alpha$ lines and one He~II line) have only UVIT coverage, another three (C~III], Al~III and He II lines) have only \hst{} coverage while the two C~IV lines are covered in both the UVIT and HST spectral range. In our subsequent spectral analysis, we have kept the emission line centroids fixed at their best-fit values. The emission lines we identified from the \hst{} spectrum of Mrk~813 are the same ones as reported by \cite{2002ApJS..143..257K}.

In the case of RBS~688, we found four emission lines that are listed in Table~\ref{Table:emission_lines}. 
We found a broad and a narrow component for the $Ly~\alpha$ line. The narrow component was unresolved, with a $90\%$ upper limit on the line width $\sigma$ at $1.6 \times 10^{-4} \kev$. Therefore, we fixed the $\sigma$ at a lower value of $10^{-5}\kev$.

To investigate the nature of intrinsic continuum emission, we replaced the power-law component by the accretion disk model {\sc diskbb} \citep{1984PASJ...36..741M,1986ApJ...308..635M}. This model describes the continuum as multi-temperature black body emission from a geometrically thin and optically thick non-relativistic accretion disk. The {\sc diskbb} model provided a satisfactory fit to both the Mrk~813  ($\chi^2/dof=1965.6/1824$), and the RBS~688 spectral data ($\chi^2/dof=160.8/157$). The fit indicated a truncated disk with $r_{in} \sim 70-100r_g$ for Mrk~813, while the inner disk appears to extend close to the innermost circular orbit for a Schwarzschild black hole in the case of RBS~688.
We further tested models for relativistic accretion disk, and replaced the {\sc diskbb} component with a fully relativistic accretion disk model {\sc zkerrbb} \citep{Li_2005}. This is a multi-temperature blackbody model to predict the continuum emission from a geometrically thin, steady-state general relativistic accretion disk around a Kerr black hole. Here we assumed a self-irradiated (by keeping the parameter `rflag' fixed at 1) relativistic accretion disk with zero torque at the inner boundary of the disk (parameter `$\eta$' = 0 ). We have not considered limb-darkening (`lflag' = -1). The parameters of this model are: mass of the central black hole $M_{BH}$ in units of solar mass, the spin parameter for the black hole $a$, inclination angle of the disk $i$ in units of degrees, the mass accretion rate $\dot{M_d}$ in units of $M_\odot~year^{-1}$, spectral hardening factor $f_{col}$ and the redshift of the source $z$. 
For both the sources, the best-fit parameters of {\sc diskbb} model indicate an inner-disk temperature of $\sim 3\times10^4$ K. Since below $3 \times 10^4$ K electron scattering becomes less important and color correction factor $f_{col} \rightarrow 1$ \citep{2011ApJ...728...98D}, we kept $f_{col}$ fixed at 1 for both the sources. It was not possible to constrain the inclination angle and the spin parameter, therefore, 
we fixed the inclination angle at two different values, $0^\circ$ and $45^\circ$ and found the best-fit values of $\dot{M_d}$ and $a$. We list the best-fit parameters in Table~\ref{Table:UV_model_pars}. We note that the {\sc zkerrbb} model provided a poorer fit ($\chi^2/dof = 2014.8/1825$) compared to the {\sc diskbb} model ($\chi^2/dof = 1965.6/1824$) in the case of Mrk~813. Also, when the spin parameter was left free, the best-fit value was $a=-0.99$ for both the inclination angles. We found a $90 \%$ upper limit on $a$  at $-0.84$ and $-0.88$ for inclination angle of $0^\circ$ and $45^\circ$, respectively. Therefore, we kept the spin parameter fixed at $a = -0.99$. This implies that the accretion disk of Mrk~813 is indeed truncated as suggested by the {\sc diskbb} model.

\subsection{X-ray Spectral Analysis}\label{subsection:Xray_analysis}
We began the X-ray spectral fitting with a simple power law ({\sc zpowerlw } model) multiplied by the Galactic absorption component {\sc tbabs} \citep{2000ApJ...542..914W}.
We fixed the absorption column at the Galactic value (see Table~\ref{Table:src_table}) for both the AGN.  Initially, we fitted this model in the $2-7\kev$ band, and then extended the best-fit power-law model down to $0.7\kev$.  
The fit residuals showed a clear soft X-ray excess in the case of Mrk~813, but no such excess was obvious in the case of RBS~688. For Mrk~813, adding a {\sc zbbody} component to the power-law model improved the $\chi^2$ by 23.3 for two additional parameters, thus with a final $\chi^2$ of 122.2 for 141 dof. For the full 0.7-7 keV band, we obtained the best-fit power-law photon index of $\Gamma = 1.7 \pm 0.2$ and a blackbody temperature of $kT = 0.10_{-0.03}^{+0.04} \kev $. 

The lack of a soft X-ray excess component in the case of RBS~688 could be due to either the poor quality of our SXT data or the intrinsic weakness of this component. Adding a {\sc zbbody} component to the power-law model did not improve the fit. Fixing the temperature at $kT=0.1\kev$, we found a $3\sigma$ upper limit to the soft X-ray excess emission of $1.7 \times 10^{-13}\rm{~ergs}~cm^{-2} s^{-1}$, which is only $\sim 10\%$ of the power-law component in the $0.7-2\kev$ band. The soft excess component contributes $\sim 17 \%$ in the case of Mrk~813. Thus, we conclude that the soft X-ray emission from RBS~688 is very weak or absent altogether.

\subsection{UV-X-ray Joint Spectral Analysis}\label{subsection:joint_analysis}
The observed UV continuum can deviate from that expected from a full accretion disk, as we have seen in section \ref{subsection:UV_analysis} in the case of Mrk~813, due to a number of factors including truncation of disk, presence of X-ray emitting regions. 
Also, the soft X-ray excess component from the innermost accretion disk can alter the emission from a standard disk. Further, the Comptonisation of disk photons in a warm or hot corona can change the shape of the disk continuum depending on the geometry of the disk/corona. To investigate these effects,  
we performed joint broadband spectral analysis of the UV (UVIT and/or HST data) and X-ray (SXT) data. 
We used the {\sc redden} component to account for the Galactic reddening in the UV band, and {\sc tbabs} component for the Galactic X-ray absorption.
We used Galactic E(B-V) and $N_H$ values as mentioned in the sections~\ref{subsection:UV_analysis} and \ref{subsection:Xray_analysis}.

We  fitted our broadband UV/X-ray spectral data  with three different models:
\begin{enumerate}[label=\roman*.]
    \item {\sc optxagnf} \citep{2012MNRAS.420.1848D}
    \item {\sc fagnsed} \citep{2018MNRAS.480.1247K}, and 
    \item {\sc relagn} \citep{10.1093/mnras/stad2499}
\end{enumerate}

While fitting with these models, we kept the energies of all the UV emission lines fixed at their best-fit values that were derived earlier and listed in Table~\ref{Table:emission_lines}. Although we varied the width of the emission lines, the best-fit values were within the error range of the values derived earlier (see Table~\ref{Table:emission_lines}). We therefore kept them frozen at their best-fit values for each individual model while calculating the errors on the model parameters.

The XSPEC model {\sc optxagnf} assumes a geometry where the accretion disk is a standard optically thick, geometrically thin disk which emits as a color-corrected multi-temperature black body down to a radius $r_{cor}$. Below this radius, an optically thick, warm inner disk produces soft X-ray excess emission while an optically thin hot corona produces broadband X-ray power-law emission. In this model, we fixed the central black hole mass $M_{BH}$ and the redshift of each source listed in Table~\ref{Table:src_table}. We fixed the distance to each source at the comoving distance calculated using the NASA/IPAC cosmology calculator \citep{2006pasp..118.1711w}. We also fixed the outer radius of the accretion disk at $ 10^5 r_g$ and the normalisation at 1. We varied the rest of the model parameters. These parameters are the Eddington ratio ($L/L_{Edd}$), black hole spin parameter $a_*$, $r_{cor}$, temperature $kT_e$ and optical depth $\tau$ of the soft Comptonising component, photon index ($\Gamma$) of the X-ray power-law component  (for which temperature was fixed at 100 keV), and the fraction $f_{pl}$ of the power below $r_{cor}$ that is emitted in the hard Comptonized component. We list the best-fit parameters in Table~\ref{Table:UV_Xray_joint_model_pars}.

The model {\sc fagnsed} and its relativistic version {\sc relagn} 
are other modified accretion disk models. The {\sc fagnsed} model assumes an accretion disk that emits as a radially stratified multi-temperature blackbody from $r_{w}$ to $r_{out}$ and an optically thick warm Comptonized disk from $r_{h}$ to $r_{w}$ that produces soft X-ray excess component. Below $r_{h}$, down to the last stable orbit $r_{ISCO}$ around the central black hole, thermal Comptonisation in a hot, optically thin corona gives rise to the broadband X-ray emission. The {\sc relagn} model is the fully relativistic version of the {\sc fagnsed} model. The relevant model parameters in both the models are mass of the central black hole $M_{BH}$, redshift ($z$), the comoving distance $d$ in Mpc, $\log{\dot{m}}$ where $\dot{m}$ is the mass accretion rate $\dot{M}$ relative to the Eddington accretion rate $\dot{M}_{Edd}$, dimensionless black hole spin parameter $a_*$, cosine of the inclination angle $i$ of the warm comptonizing medium and the outer disk, electron temperature $kT_h$ and spectral index $\Gamma_{h}$ of the hot Comptonising component, electron temperature $kT_{w}$ and spectral index $\Gamma_{w}$ of the warm Comptonizing component, and the outer radius of the disk in units of $r_g$. We fixed the temperature of the hot Comptonizing component $kT_h$ to 100 keV and the model normalisation to 1. For both the targets, we let the model calculate the color correction factor $f_{col}$ at each radius of the standard accretion disk by fixing the parameter $f_{col}$ at -1.

\subsection{Mrk~813}
In the {\sc optagnf} model, we fixed the spin parameter to zero, and the temperature of the hot Comptonisation component $kT_{e,h}$ to $100\kev$. The best-fit model resulted in $\chi^2/dof = 2103.3/1964$ with the warm corona  $kT_{e,w} = 0.065_{-0.005}^{+0.006}\kev$, $\tau > 70$ and $r_{cor} > 70 r_g$, and the hot corona producing a Comptonised power law with $\Gamma=1.7_{-0.1}^{+0.1}$. The best parameters, listed in Table~\ref{Table:UV_Xray_joint_model_pars}, are typical of AGN with high accretion rates.

While fitting the {\sc fagnsed} and {\sc relagn} models, we fixed the cosine of the inclination angle at 0.99, which corresponds to an inclination angle of $\sim 0^\circ$. We have also fixed the dimensionless spin parameter to zero (since these models do not allow negative spin cases) as inferred from {\sc zkerrbb}. While fitting, we also noticed that the spectral index for the warm Comptonization component, $\Gamma_w$, was always pegged at its lower limit, so we fixed it at its lower limit of 2.0, and that helped us to constrain the other parameters of the model. We have also tested the {\sc fagnsed} model for four different inclination angles, $0^\circ, 25^\circ, 35^\circ$ and $45^\circ$. We found that the fit statistics did not change substantially for different inclination angles, only the radius of the warm componizing disk and the hot corona changed slightly depending on the inclination angles, and the Eddington ratio varied in the range $ 0.29 - 0.50$. Similarly, with the {\sc optxagnf} model, we fixed the spin parameter, $ a _*$, at 0. The spectral index of the hard X-ray component was similar in all three models, while {\sc optxagnf} model predicted a slightly lesser electron temperature for the soft X-ray component at $0.069_{-0.006}^{+0.007} \kev$ than $\sim 0.1 \kev$ predicted by {\sc fagnsed} and {\sc relagn} models. Among the three models, {\sc optxagnf} predicted a higher Eddington ratio ($L/L_{Edd}$) of $\sim 0.7$, followed by $\sim 0.29$ and $\sim 0.26 $ predicted by {\sc fagnsed} and {\sc relagn}, respectively.  A full list of all the best-fit parameters for the three models is given in Table~\ref{Table:UV_Xray_joint_model_pars}.

\begin{table*}
    \caption{The UV emission line parameters of Mrk~813 and RBS~688. The t lines ($f_{\rm{line}}$) is expressed in units of $\rm{photons}~\cm^{-2} \s^{-1}$.}
    \label{Table:emission_lines}
    \centering
    \begin{tabular}{c c c c c}
        \hline\hline
        Emission line & Parameter & \multicolumn{2}{c}{Mrk 813} & RBS 688 \\
        \cmidrule(lr){3-4}
         & &  UVIT & HST & UVIT \\
         \hline
         Ly $\alpha$ & $\lambda$ (\AA) & -- & 1224 & 1220 \\
         & $v_\sigma$ (km/s) & -- & $11812_{-1787}^{2103}$ & $6129_{-342}^{+683}$ \\
         & $f_{\rm{line}}$  & -- & $0.20_{-0.03}^{+0.03}$ & $0.18_{-0.01}^{+0.02}$\\
         \cmidrule(lr){2-5}
         & $\lambda$ (\AA) & -- & 1216 & 1220 \\
         & $v_\sigma$ (km/s) & -- & --  & -- \\
         & $f_{\rm{line}}$ & -- & $0.11_{-0.02}^{+0.02}$ & $0.028_{-0.004}^{+0.007}$\\

         \hline
         
         C IV & $\lambda$ (\AA) & 1546 & 1546 & 1515 \\
         & $v_\sigma$ (km/s) & $4460_{-389}^{+511}$  & $4460_{-389}^{+511}$ & $1947_{-1058}^{+1407}$\\
         & $f_{\rm{line}}$ & $0.13_{-0.02}^{+0.02}$ & $0.13_{-0.02}^{+0.02}$ & $0.05_{-0.02}^{+0.03}$\\
         \cmidrule(lr){2-5}
         & $\lambda$ (\AA) & 1550 & 1550 & --\\
         & $v_\sigma$ (km/s) & $1455_{-458}^{+497}$  & $1455_{-458}^{+497}$ & --\\
         & $f_{\rm{line}}$ & $0.03_{-0.01}^{+0.02}$ & $0.03_{-0.01}^{+0.02}$ & -- \\

         \hline
         
         C III] & $\lambda$ (\AA) & 1903 & -- & -- \\
         & $v_\sigma$ (km/s) & $3315_{-264}^{+288}$  & -- & -- \\
         & $f_{\rm{line}}$ & $0.04_{-0.004}^{+0.004}$ & -- & -- \\
         \hline
         Al III & $\lambda$ (\AA) & 1835 & -- & -- \\
         & $v_\sigma$ (km/s) & $3107_{-1804}^{+1767}$ & -- & -- \\
         & $f_{\rm{line}}$ & $0.006_{-0.003}^{+0.003}$ & -- & -- \\

         \hline
         
         He II & $\lambda$ (\AA) & 1623 & -- & -- \\
         & $v_\sigma$ (km/s) & $5825_{-758}^{842}$ & -- & -- \\
         & $f_{\rm{line}}$ & $0.05_{-0.009}^{+0.01}$ & -- & -- \\

         \hline
         
         Si IV & $\lambda$ (\AA) & -- & 1385 & -- \\
         & $v_\sigma$ (km/s) & -- & $5835_{-1635}^{+4533}$ & -- \\
         & $f_{\rm{line}}$ & -- & $0.05_{-0.01}^{+0.03}$ & -- \\
         \hline
         Si II & $\lambda$ (\AA) & -- & -- & 1313 \\
         & $v_\sigma$ (km/s) & -- & -- & $4269_{-2864}^{+3723}$ \\
         & $f_{\rm{line}}$ & -- & -- & $0.03_{-0.01}^{+0.02}$ \\
         \hline
         
    \end{tabular}
\end{table*}

\begin{figure*}[ht!]
\gridline{\fig{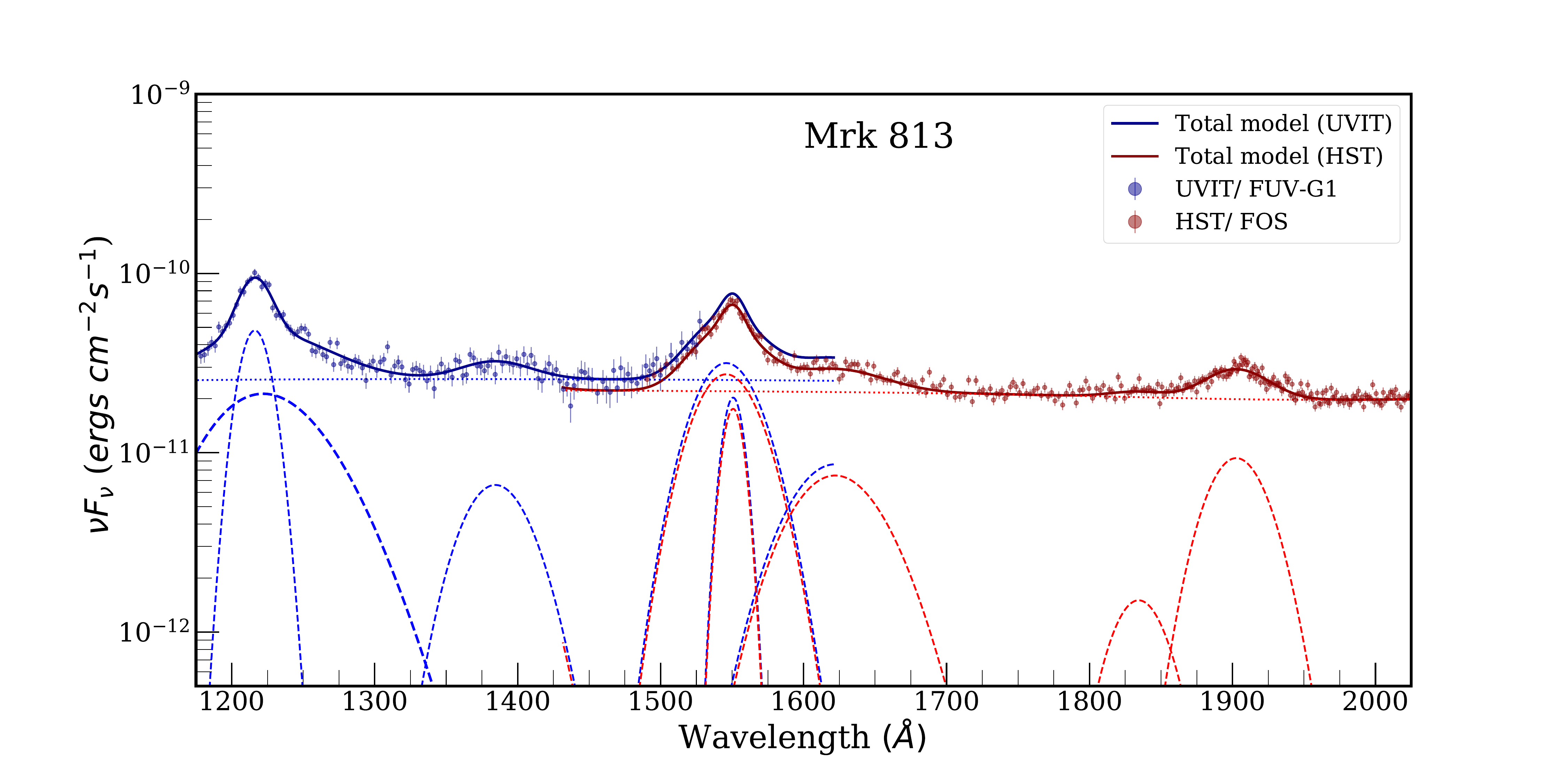}{\textwidth}{}}
\gridline{\fig{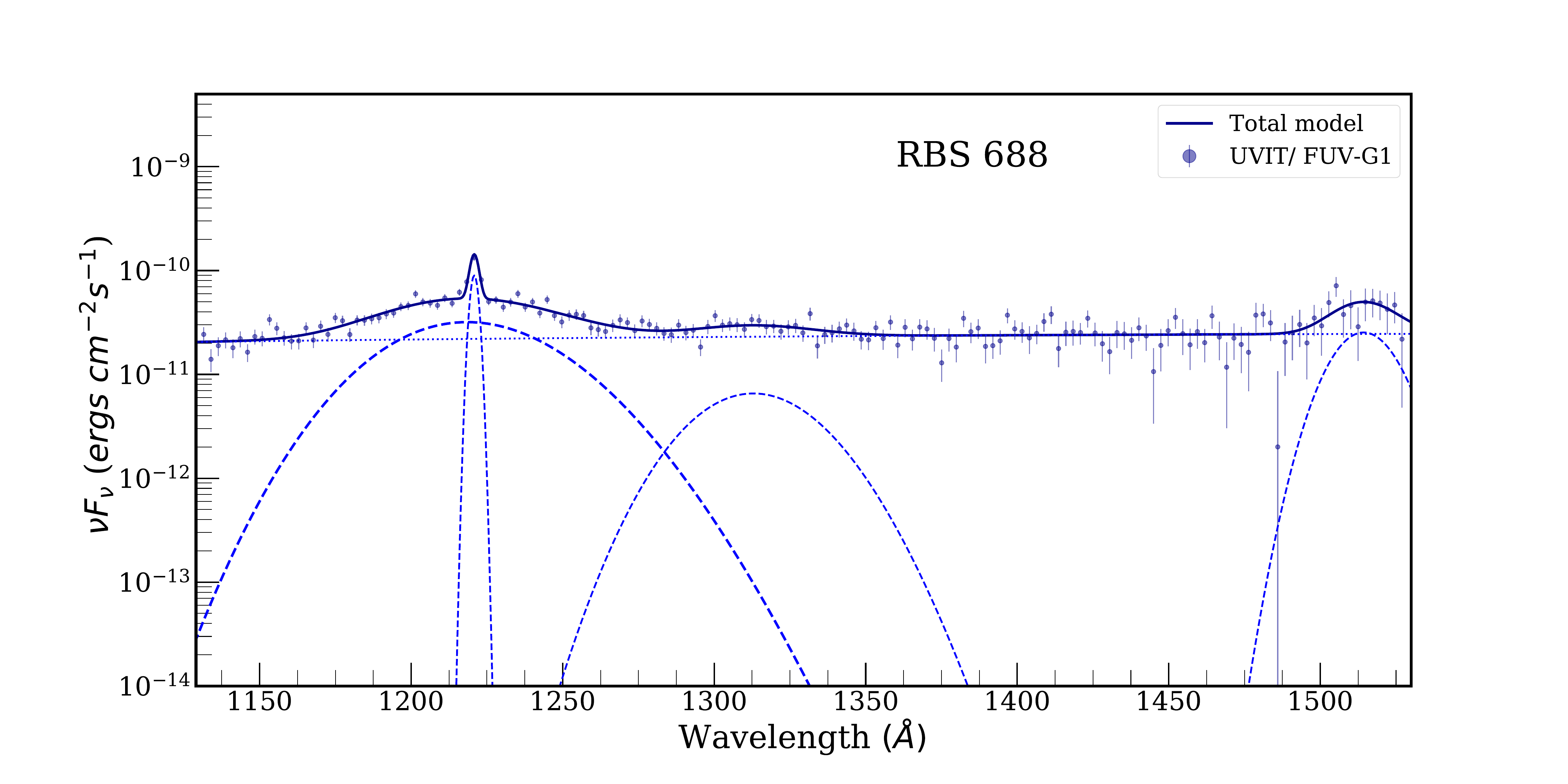}{\textwidth}{}}
\caption{The UV emission Line spectra of Mrk~813 (upper panel) and RBS~688 (lower panel). The X-axis represents rest-frame wavelengths. UVIT/FUV-G1 spectral data are marked in dark blue, while HST data are marked in dark red. For each dataset, the total model is shown as dark (dark-blue for UVIT and dark-red for HST) solid lines, while the power-law continuum is shown as a dotted line, and the emission lines are shown as dashed lines.}
\end{figure*}

\begin{table*}
\caption{The best-fit model parameters derived from  the spectral fitting of the UVIT/FUV and \hst{}/FOS data.}
\begin{center}
\begin{tabular}{l l c c}
\hline \hline
Model & Parameter & Mrk~813 & RBS~688 \\
\hline
\multirow{5}{8em}{\sc diskbb} & $T_{\text{in}}$ ($10^4$ K) & $3.2_{-0.2}^{+0.3}$ & $3.6_{-0.5}^{+1}$\\
& Norm  & $5_{-1}^{+2}\times10^{10}$ & $4_{-2}^{+6}\times10^{10}$\\
& $R_{\text{in}}$ ($ r_{g}$) (for i = $0^{\circ}$) & $77_{-6}^{+14}$ & $9_{-3}^{+5}$ \\
& $R_{\text{in}}$ ($r_{g}$) (for i = $45^{\circ}$) & $91_{-7}^{+17}$ & $11_{-4}^{+6}$\\
& $\chi^2/dof$ & 1965.6/1824 & 160.8/157 \\
\hline
\multirow{3}{8em}{\sc zkerrbb ($i = 0^\circ$)~} &$a_*$ & $ < -0.84$ & 0.98 (f)\\
& $\dot{M} (M_\odot/yr )$ & $0.82_{-0.03}^{+0.03}$ & $0.48_{-0.03}^{+0.03}$ \\
& $\chi^2/dof$ & 2014.8/1825  & 160.8/158 \\
\hline
\multirow{3}{8em}{\sc zkerrbb ($ i = 45^\circ$)} &$a_*$ & $< -0.88$ & $0.7_{-0.7}^{+0.3}$ \\
& $\Dot{M} (M_\odot/yr )$ & $1.17_{-0.04}^{+0.04}$ &  $1.0_{-0.5}^{+0.8}$\\
& $\chi^2/dof$ & 2036.4/1825 &  160.8/157 \\
\hline
\end{tabular}

\label{Table:UV_model_pars}
\end{center}
\end{table*}

\begin{table*}
\caption{The best-fit parameters derived from the joint spectral analysis of UVIT/FUV, \hst{}/FOS and SXT X-ray data with the {\sc optxagnf}, {\sc fagnsed,} and {\sc relagn} models.}
\label{Table:UV_Xray_joint_model_pars}
\begin{center}
\begin{tabular}{c | c |c c c }
\hline \hline
Source  & Parameter & {\sc optxagnf} & {\sc fagnsed} & {\sc relagn} \\
\hline
\multirow{15}{5em}{Mrk 813} & $\log{L/L_{Edd}}$ & $-0.15_{-0.06}^{+0.03}$ & $-0.54_{-0.07}^{+0.1}$ & $-0.59_{-0.06}^{+0.09}$ \\
& $a_{*}$ & 0 & 0 & 0 \\
& $\cos{i}$ & -- & 0.99 & 0.99\\
& $kT_{e,h}$ (keV) & 100  & 100 & 100\\
& $kT_{e,w}$ (keV) & $0.065_{-0.005}^{+0.006}$ & $0.09_{-0.01}^{+0.04}$ & $0.11_{-0.02}^{+0.06}$\\
& $\Gamma_h$ & $1.8_{-0.1}^{+0.1}$ & $1.7_{-0.2}^{+0.1}$ & $1.7_{-0.2}^{+0.1}$\\
& $\Gamma_w$ & -- & $2.0$ & $2.0$ \\
& $\tau$ & $ > 70 $ & -- & --\\
& $r_h$ ($r_g$) & -- & $16_{-4}^{+18}$ & $23_{-6}^{+18}$\\
& $r_{cor}$ or $r_w$ ($r_g$) & $ > 70 $ & $76_{-21}^{+38}$ & $62_{-17}^{+26}$ \\
& $f_{pl}$ & $0.12_{-0.02}^{+0.05}$ & -- & -- \\
& $\log{r_{out}}$ & 5 & 5 & 5 \\
& $h_{max}$ & -- & $r_h$ & $r_h$\\
& $\chi^2$ / d.o.f & 2103.3/1964 & 2107.4/1965 & 2106.8/1965 \\
\hline
\multirow{15}{5em}{RBS 688} &  $\log{L/L_{Edd}}$ & $-1.10_{-0.04}^{+0.03}$ & $-1.27_{-0.04}^{+0.03}$ & $-1.16_{-0.03}^{+0.03}$\\
& $a_{*}$ &  0.99 & 0.99 & 0.99 \\
& $\cos{i}$ & -- & 0.7 & 0.7\\
& $kT_{e,h}$ (keV) & 100  & 100 & 100\\
& $kT_{e,w}$ (keV) & 0.1 & 0.1 & 0.1\\
& $\Gamma_h$ & $2.1_{-0.1}^{+0.1}$ & $2.0_{-0.1}^{+0.1}$ & $2.0_{-0.1}^{+0.1}$\\
& $\Gamma_w$ & -- & 2.0 & 2.0 \\
& $\tau$ & 1.0 & -- & -- \\
& $r_h$ ($r_g$) & --  & $2.01_{-0.07}^{+0.07}$  & $3.7_{-0.1}^{+0.2}$\\
& $r_{cor}$ or $r_w$ ($r_g$) & $2.00_{-0.04}^{+0.05}$ & $r_h$ & $r_h$\\
& $f_{pl}$ & 1.0 & -- & -- \\
& log $r_{out}$ & 5 & 5 & 5 \\
& $h_{max}$ & -- & 10 & 10\\
& $\chi^2/dof$ & 319.84/315  & 319.84/315  & 320.24/315 \\
\hline
\end{tabular}
\end{center}
\end{table*}

\begin{figure*}
\gridline{\fig{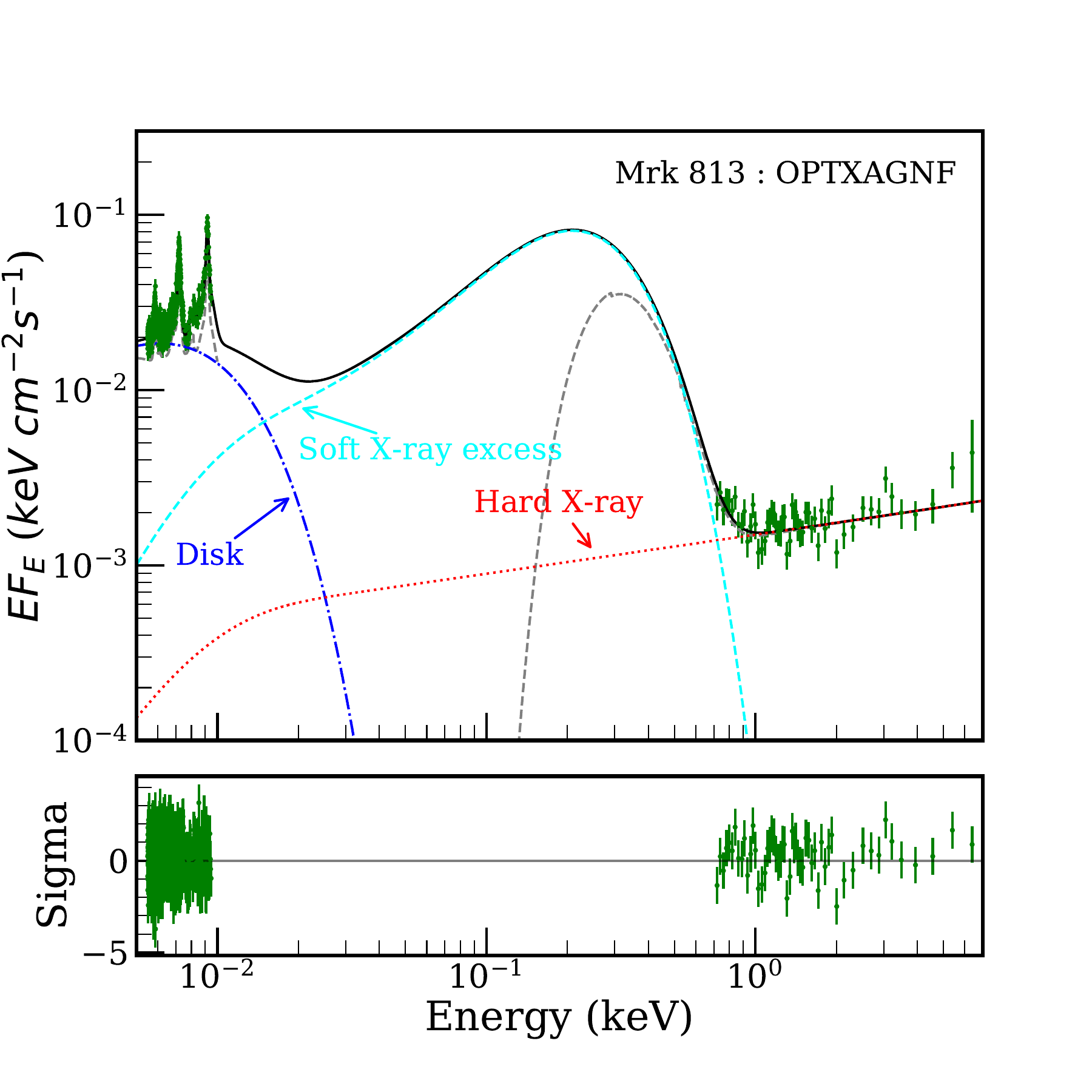}{0.5\textwidth}{}}
\gridline{\fig{mrk813_fagnsed_SED_plot.pdf}{0.5\textwidth}{}
            \fig{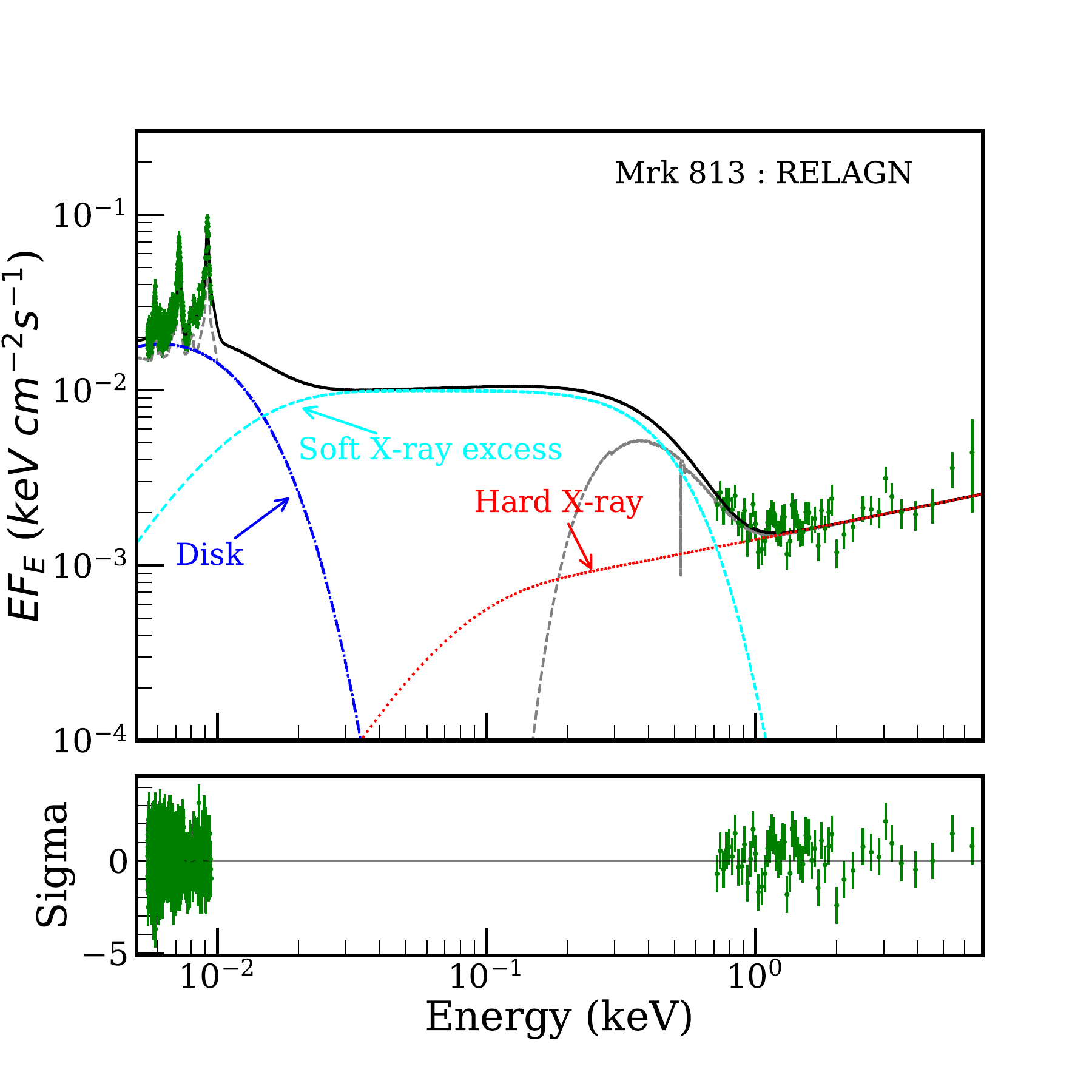}{0.5\textwidth}{}}
\caption{Broadband spectral data on Mrk 813 fitted with different models. The observed data are fitted with absorbed broadband SEDs, which are shown as dashed grey curves. 
Also shown are the absorption-corrected data (green) and full models. The absorption-corrected individual model components -- the disk continuum (dot-dashed blue curve), hard X-ray component (dotted red curve) and soft X-ray excess component (dashed cyan curve) are also shown.
\label{fig:Mrk813_broadband_SED}}
\end{figure*}

\subsection{RBS~688}
As mentioned in section~\ref{subsection:Xray_analysis}, RBS~688 lacked strong soft X-ray excess emission. This component contributes less than $9\%$ ($3\sigma$ upper limit) in the $0.7-2\kev$ band. 
Therefore, while fitting with the {\sc optxagnf} model, we fixed the parameter $f_{pl}$, which measures the fraction of power below $r_{cor}$ that is emitted in the hard X-ray component, at 1.0. This indicates, below $r_{cor}$, power is completely emitted in the form of broadband X-ray power law. We fixed the other two parameters related to the soft X-ray component, namely the electron temperature $kT_{w}$ and the optical depth $\tau$ of the warm comptonizing medium  at 0.1 and 1.0, respectively. However, these parameters would not affect the fit since the contribution of soft X-ray component in the total emitted energy was effectively made zero. Initially, we kept the spin parameter $a_*$ free, but could not constrain it and only found the lower limit at 0.99, so we fixed it at 0.99 while calculating the error on other parameters. The best-fit values of all relevant parameters are mentioned in Table~\ref{Table:UV_Xray_joint_model_pars}.

\begin{figure*}
\gridline{\fig{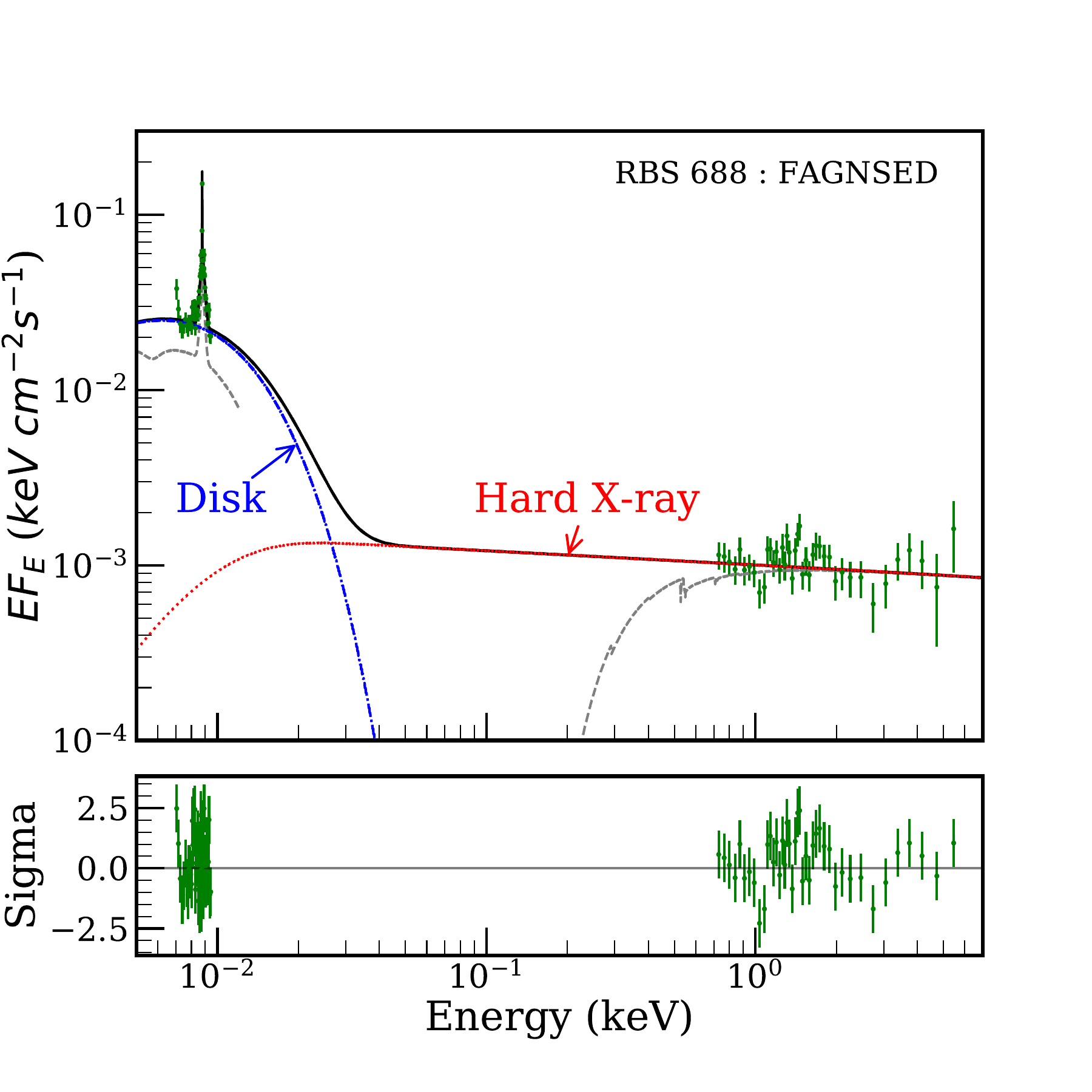}{0.5\textwidth}{}
          \fig{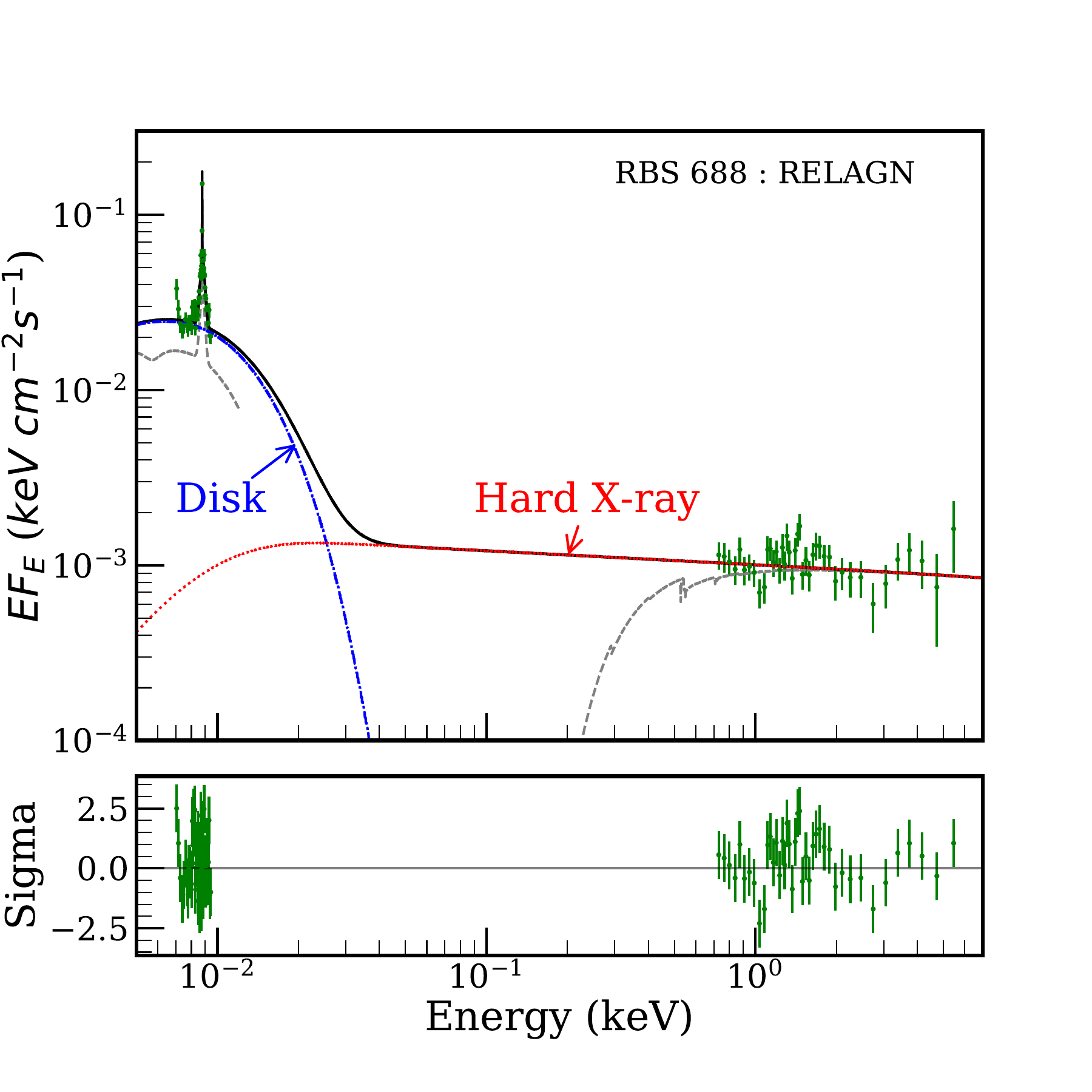}{0.5\textwidth}{}}
\gridline{\fig{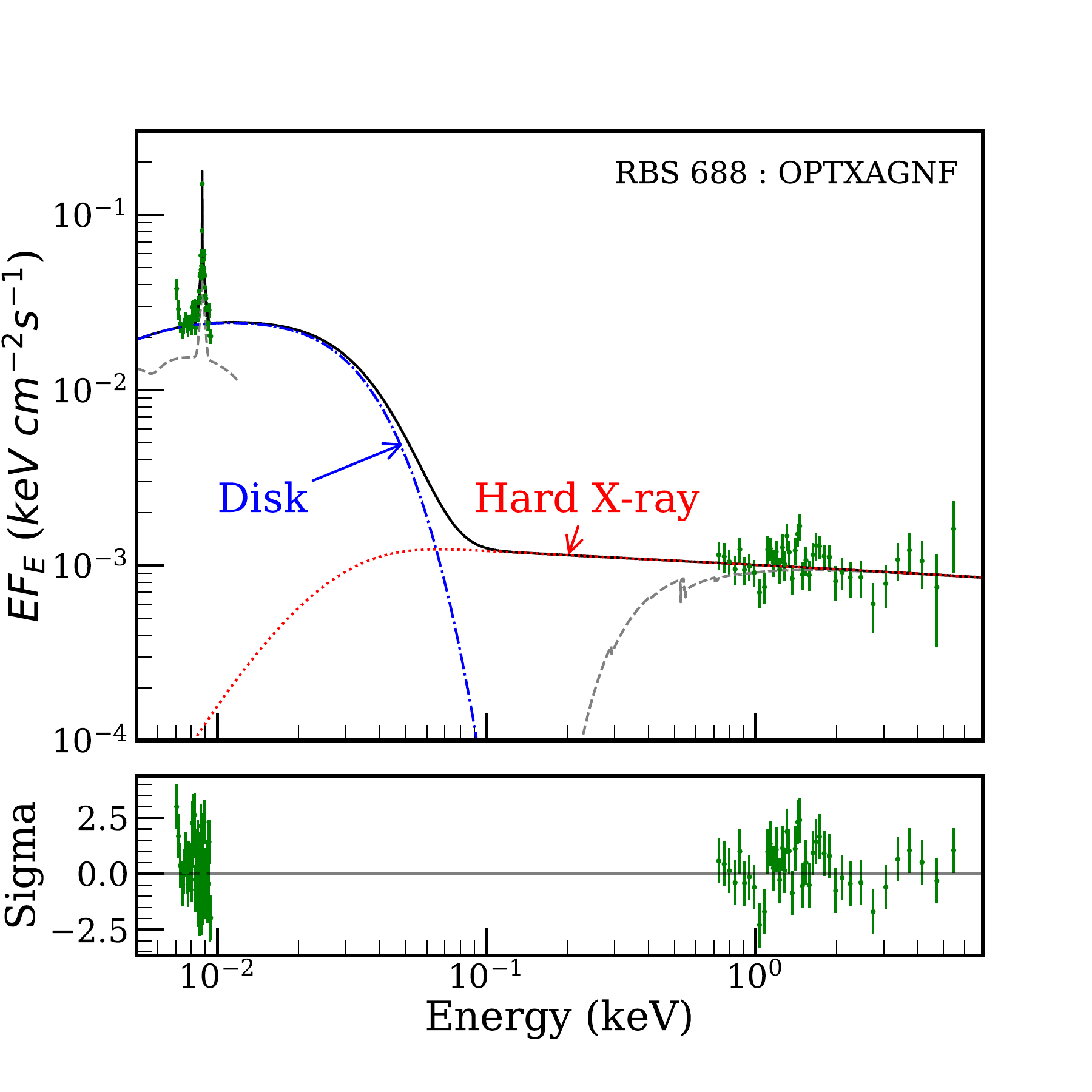}{0.5\textwidth}{}}
\caption{The broadband UV/X-ray spectral data and the best-fitting models for RBS~688. The UVIT/FUV and SXT spectral data, the best-fitting models and the data-to-model residuals in terms of Sigma (i.e., data--model/errors).   
The observed data are fitted with absorbed broadband SEDs, which are shown as dashed grey curves. Also shown are the absorption-corrected data (green) and full models and the continuum model components (accretion disk continuum in dot-dashed blue, Thermal Comptonised component from hot corona emission as dotted red line).
\label{fig:RBS688_broadband_SED}}
\end{figure*} 

As in the case of {\sc optxagnf} model, we set the parameters of our broadband spectral models {\sc fagnsed} and {\sc relagn} to remove the soft X-ray excess component. We tied the 
radius of the warm Comptonising medium $ r_w$ to the radius of the hot Comptonising medium $r_h$ in the {\sc fagnsed} and {\sc relagn}. This corresponds to a geometry where the color-corrected standard disk extends down to $r_h$ and below this radius, a hot Comptonising corona produces the X-ray power-law component.

While the {\sc fagnsed} and {\sc relagn} models, we could only find the $90\%$ lower limit on spin parameter at $a_*=0.8$.  Fixing the spin parameter at $a_* = 0.99$ resulted in  better constraints on $r_h$ and log($\dot{M}/ {\dot{M}_{Edd}}$). Both {\sc fagnsed} and {\sc relagn} predicted that the standard accretion disk extends very close to the ISCO for a maximally rotating Kerr black hole. The {\sc fagnsed} model predicts a slightly lower Eddington ratio at $\sim 0.054$ compared to $\sim 0.069$ predicted by the {\sc relagn} model. 

Testing for different inclination angles at $i = 0^\circ, 10^\circ, 25^\circ, 35^\circ$, and $45^\circ$, the {\sc fagnsed} model did not result in any significant difference in $\chi^2$, while the {\sc relagn} model showed improvement with $i= 45^\circ$ ($\chi^2 /d.o.f. \sim 325.2/315)$ than that with $i=0^\circ$ ($\chi^2/ d.o.f \sim 320.0 /315$). We found the best-fit Eddington ratio in the range of 0.05 (for $i = 0^\circ$) to 0.07 (for $i = 45^\circ$), and the size of the hot corona $r_h$ at 6.9, 6.8, 5.7, 4.6 and 3.7 $r_g$, for $i = 0^\circ, 10^\circ, 25^\circ, 35^\circ$ and $45^\circ$, respectively. The Table~\ref{Table:UV_Xray_joint_model_pars} lists the best-fit parameters corresponding to the lowest $\chi^2$ among these five cases, i.e. the one corresponding to the inclination angle of $45^\circ$.

Thus, in the case of RBS~688, all three models resulted in a similar fit, although {\sc optxagnf} predicts the highest Eddington ratio at $\sim 0.079$. A full list of all the best-fit parameters for all three models is mentioned in Table~\ref{Table:UV_Xray_joint_model_pars}, and the fitted broadband spectra are shown in Figure~\ref{fig:RBS688_broadband_SED}.

\section{Results and Discussion}\label{sect:result}
We analysed simultaneous far UV and X-ray spectral data acquired with the \astrosat{} observations of two AGNs, Mrk~813 and RBS~688. We additionally used the \hst/FOS spectrum of Mrk~813 that has similar flux as our \astrosat{}/UVIT spectrum (see Fig.~\ref{fig:UV_spectrum}). These broadband UV/X-ray spectral data allowed us to probe the inner accretion disk regions in two massive AGNs.

\subsection{Mrk~813}
The UVIT/FUV and \hst/FOS far UV spectra of Mrk~813 agree well with a simple power-law continuum, 
and emission lines from the BLR/NLR, all affected by the Galactic reddening only. The best-fit photon-index $\Gamma = 1.9_{-0.1}^{+0.2}$, corresponding to a spectral index of $\alpha = 1 - \Gamma = -0.9_{-0.1}^{+0.2}$, is steeper than that predicted for the standard disk model ($f_\nu \propto \nu^{1/3}$). This indicates that there is a deficit of emission at shorter wavelengths when compared to a standard disk emission. Such a deficit may arise from internal reddening due to the host galaxy. But as explained in section \ref{subsection:UV_analysis}, we did not find any evidence of internal reddening. Another possible reason for the short-wavelength deficit could be due to the contamination from the host galaxy emission. However, Mrk~813 is a bright Seyfert 1 galaxy at a redshift of $z=0.111$. However, the AGN appears as a point source in the UV images without any evidence for extended emission from the host galaxy, which indicates that the host galaxy does not contribute much to the observed spectrum. 
Also, the superior spatial resolution of \hst{} makes the FOS spectrum nearly free from the host galaxy contamination, as we do not see stellar absorption features expected from the host galaxy.

A deviation from the $f_{\nu} \propto \nu^{1/3}$ spectral shape may occur if the UV spectral bandpass includes the peak or extends to wavelengths shorter than the peak of the disk emission. 
For a standard accretion disk around a SMBH of mass  $M_{BH} \sim 9\times10^7 M_{\odot}$  accreting at $\sim 1{M_{\odot}~yr^{-1}}$, the emission is expected to peak near $\sim 1000{\rm~\AA}$. Thus, our UVIT/FUV and \hst{}/FOS bands cover longward of the expected peak emission from a full standard disk. Therefore, the observed steeper UV continuum of Mrk~813, relative to that of a standard accretion disk, indicates that its actual accretion disk deviates in some way from the standard model.
The standard disk model {\sc diskbb} fitted to the UV spectra of Mrk~813 provided a temperature of $T_{in} = 3.2_{-0.2}^{+0.3} \times 10^4{\rm~K}$ at the inner edge. 
Using the best-fit value of the {\sc diskbb} normalisation, we estimated the innermost radius  at $r_{in} = 77_{-6}^{+14}~r_g$ and $91_{-7}^{+17}~r_g$ at inclination angles of $0^\circ$ and $45^\circ$, respectively. This suggests that the disk is truncated in the inner regions.  

With the fully relativistic accretion disk model ({\sc zkerrbb}), we were not able to constrain the spin parameter $a_*$. We only found $90\%$ upper limits of $a_* \le -0.84$ and $\le -0.88$ for $0^\circ$ and $45^\circ$ inclination angle, respectively. We derived an accretion rate in the range of $0.8$ to $1.2{\rm~M_{\odot}~yr^{-1}}$ for inclination angle $0$ to $45^{\circ}$ (see Table~\ref{Table:UV_model_pars}). For the black hole mass of $M_{BH} = 9 \times 10^7~M_{\odot}$, this corresponds to an Eddington fraction of $\dot{m}\sim 0.4-0.6$.
The upper limit on the spin parameter of $a_* \le -0.84$ suggests that the central black hole of Mrk~813 has a negative spin, or most likely the inner disk is truncated at a radius $r_{in} \gtrsim 9r_{rg}$ as inferred in the case of {\sc diskbb} model.  
The replacement of the {\sc diskbb} model with the {\sc zkerrbb}  worsened the fit statistics (see Table~\ref{Table:UV_model_pars} as the {\sc zkerrbb} model does now allow for disk truncation. This further suggests that Mrk~813 does not host a full standard disk. The deficit of emission at short UV wavelengths can arise if the inner disk is modified in some way so that the innermost regions do not contribute to the UV emission, or if there is no material in the inner region i.e., the disk is truely truncated. The first possibility is explored in the case of Comptonized accretion disk models, where the innermost regions of the disk are at a higher temperature in the form of an optically thick warm corona, which produces the soft X-ray excess component below $\sim 1\kev$. We explored this possibility by performing joint spectral fitting of the UV and X-ray data with three different models, namely {\sc fagnsed, relagn} and {\sc optxagnf} (see Table~\ref{Table:UV_Xray_joint_model_pars}). 
These models suggest that Mrk~813 does not host a full standard accretion disk down to the innermost stable orbit.  Most likely, the innermost disk regions exist only in the form of a warm, Comptonized disk, giving rise to the soft X-ray excess component.

\subsection{RBS~688}
The UVIT/FUV spectrum of RBS~688 is also consistent with a simple power-law continuum and Gaussian emission lines. 
 The power-law photon index of $\Gamma=2.3 \pm 0.5$  ($\alpha = -1.3\pm 0.5$) is much steeper than the $\nu^{1/3}$ shape expected for the optical/UV continuum from a standard disk. In the case of RBS~688,  a SMBH of mass $M_{BH} \sim 9.5\times10^8M_{\odot}$ accreting at a rate $\dot{M}\sim 1{\rm~M_{\odot}~yr^{-1}}$ ($L/L_{Edd} \sim 0.05$), a standard disk is expected to peak near $\sim 2000{\rm~\AA}$. Our FUV band lies shortward of this peak, i.e., in the exponentially decaying part of the disk emission. Hence, the steep observed continuum is not necessarily a signature of departure from a standard disk. 
Indeed, the multi-color disk blackbody  {\sc diskbb} model provided a good fit with 
a temperature of $3.6_{-0.5}^{+1} \times 10^4$~K at the innermost edge of the disk. The {\sc diskbb} model also suggested the innermost radius to be at $9_{-3}^{+5} r_g$ for a face-on disk (see Table~\ref{Table:UV_model_pars}). Thus, the {\sc diskbb} model suggests that the disk extends close to the ISCO, where relativistic effects may be important.
 
The relativistic disk model {\sc zkerrbb} also described the UVIT data well, but we could not constrain the inclination and the spin parameter. 
For $i=0^\circ$, the $90\%$ lower limit on the spin parameter of $a_* \ge 0.61$. We obtained a mass accretion rate of $\dot{M} = 0.48\pm0.03{\rm~M_{\odot}~yr^{-1}}$ for $i=0^\circ$ and $a_* = 0.98$. For  i=$45^\circ$ inclination, the spin parameter is constrained to $a_*=0.7_{-0.7}^{+0.3}$ and the accretion rate is  $\dot{M} = 1.0_{-0.5}^{+0.8}~M_\odot \rm{yr}^{-1}$ (see Table~\ref{Table:UV_model_pars}). The mass accretion rate corresponds to a low Eddington ratio, $L/L_{Edd} \sim 0.02-0.05$. 

As can be seen in Table~\ref{Table:UV_Xray_joint_model_pars}, the three SED models suggest presence of full standard disk without an inner warm corona. These models also predicted a very compact corona with size $2-4r_g$, similar to the size estimates based on X-ray variability, X-ray reflection, and X-ray micro-lensing \citep[see][for a review]{2025FrASS..1130392L}.

Comparing the broadband UV/X-ray emission from Mrk~813 and RBS~688, it appears that the warm Comptonizing region responsible for the soft excess exists only at high Eddington fraction, as in the case of Mrk~813, while at low Eddington fraction, as in the case of RBS~688, the warm corona does not exist. This result is similar to what is observed in changing-look AGNs that show strong soft X-ray excess in the high flux states and no soft X-ray excess at low flux states \citep[see e.g.,][]{2022ApJ...925..101T,2022ApJ...930..117T}.

\section{Conclusions}

In this paper, we presented UV/X-ray broad broadband spectra of two bright AGN Mrk~813 and RBS~688 and investigated the inner accretion flow using accretion models. The main results of our study are as follows.
\begin{enumerate}
    \item The \astrosat{}/UVIT and \hst{}/FOS UV spectra of Mrk~813 clearly show a deficit of continuum emission compared to that expected from a standard accretion disk, while the \astrosat{}/SXT data show soft X-ray excess emission below $\sim 1\kev$ over and above a broadband X-ray power law component.
    \item The AGN in Mrk~813 is accreting at a high rate ($25-70\%$ of the Eddington rate). At this rate, its accretion disk emits like a standard disk down to only $\sim70r_g$ below which the disk appears to be in the form of a warm ($kT_e \sim 0.1\kev$) Comptonising corona.  
    \item Unlike Mrk~813, RBS~688 neither shows a UV deficit nor soft X-ray excess emission. 
    \item RBS~688 is accreting at a low rate ($\sim 6\%$ of the Eddington limit), it appears to host a full standard accretion disk without a warm corona. 
\end{enumerate}

\begin{acknowledgements}
This publication utilizes the data from Indian Space Science Data Centre (ISSDC) of the \astrosat mission of the Indian Space Research Organisation (ISRO). This publication uses UVIT data processed by the payload operations center at Indian Institute of Astrophysics (IIA). The UVIT is built in collaboration between IIA, Inter-University Centre for Astronomy and Astrophysics (IUCAA), Tata Institute of Fundamental Research(TIFR), ISRO, and the Canadian Space Agency (CSA). UVIT data were reprocessed by CCDLAB pipeline. The raw data we used for our analysis can be downloaded from \astrosat{} data archive: \url{https://astrobrowse.issdc.gov.in/astro_archive/archive/Home.jsp}. This publication used
archival FOS spectra from the HST data archive (\url{https://archive.stsci.edu/hst/search.php}) and FOS spectra from \cite{2002ApJS..143..257K} available at \url{https://hea-www.harvard.edu/~pgreen/HRCULES.html}. This research has made use of the SIMBAD database, operated at CDS, Strasbourg, France. This research has made use of the NASA/IPAC Extragalactic Database (NED), which is funded by the National Aeronautics and Space Administration and operated by the California Institute of Technology. This research has made use of data and/or software provided by the High Energy Astrophysics Science Archive Research Center (HEASARC), which is a service of the Astrophysics Science Division at NASA/GSFC. This research used various Python and Julia packages.
\end{acknowledgements}

%

\vspace{5mm}
\facilities{\astrosat, \hst}


\software{CCDLAB \citep{2017PASP..129k5002P}, XSpec \citep{1996ASPC..101...17A}, Julia \citep{Julia-2017}, SAOImage DS9 \citep{2003ASPC..295..489J}, Astropy \citep{astropy:2022}, Matplotlib \citep{Hunter:2007}.}






\bibliography{Paper_biblio}{}
\bibliographystyle{aasjournal}



\end{document}